\documentclass[pamm,a4paper,fleqn]{w-art}
\usepackage{times,cite,w-thm}
\usepackage[T1]{fontenc}
\usepackage[utf8]{inputenc}
\usepackage{graphicx}
\begin{document}

\TitleLanguage[EN]
\title[The DRESDYN project]{The DRESDYN project: planned experiments and present status}

\author{\firstname{Frank} \lastname{Stefani} \inst{1} 
\footnote{Corresponding author: e-mail \ElectronicMail{F.Stefani@hzdr.de}, 
     phone +49\,351\,260\,3069,
     fax +49\,351\,260\,2007}} 
\address[\inst{1}]{\CountryCode[DE]Helmholtz-Zentrum Dresden-Rossendorf, Bautzner Landstr. 400, 01328 Dresden, Germany}
\author{\firstname{Sven} \lastname{Eckert}\inst{1}}
\author{\firstname{Gunter} \lastname{Gerbeth}\inst{1}}
\author{\firstname{Andr\'e} \lastname{Giesecke}\inst{1}}
\author{\firstname{Thomas} \lastname{Gundrum}\inst{1}}
\author{\firstname{Dirk} \lastname{R\"abiger}\inst{1}}
\author{\firstname{Martin} \lastname{Seilmayer}\inst{1}}
\author{\firstname{Tom} \lastname{Weier}\inst{1}}

\AbstractLanguage[EN]
\begin{abstract}
The Dresden sodium facility for dynamo and thermohydraulic 
studies (DRESDYN) is a platform for large-scale liquid 
sodium experiments devoted to fundamental geo- and
astrophysical questions as well as to various applied 
problems related to the conversion and storage of energy. 
Its most ambitious part is a precession driven dynamo experiment,
comprising 8 tons of liquid sodium supposed to rotate with 
up to 10 Hz and to precess with up to 1 Hz. Another 
large-scale 
set-up is a Tayler-Couette experiment with a gap
width of 0.2 m and a height of 2 m, whose inner cylinder 
rotates with up to 20 Hz. Equipped with a coil system
for the generation of an axial field of up to 120 mT  
and two different 
axial currents through the center and the liquid sodium,
this experiment aims at studying various versions of the 
magnetorotational instability and their combinations with 
the Tayler instability. We discuss the physical
background of these two experiments and 
delineate the present status of their
technical realization. Other installations, such as 
a sodium loop  and a test stand for
In-Service-Inspection experiments 
will also be sketched.
\end{abstract}
\maketitle                   

\section{Introduction}

The large-scale infrastructure DRESDYN (DREsden Sodium facility 
for DYNamo and thermo-hydraulic studies) at Helmholtz-Zentrum 
Dresden-Rossendorf is a platform for large- and medium-sized 
liquid-sodium experiments \cite{Stefani2012}, 
partly devoted to fundamental 
problems of geo- and astrophysics, partly 
intended for 
investigations into energy-related topics, including 
various In-Service-Inspection (ISI) aspects \cite{Tenchine2010} 
of Sodium-Fast-Reactors (SFRs), 
and stability problems of 
liquid metal batteries \cite{Weber2013,Weber2014}. 
In this paper, we will
present the general structure of the DRESDYN project, 
the main experimental installations 
and their scientific background.

\section{DRESDYN - An overview}

The liquid sodium experiments will be  carried out in a
new laboratory building with approximately 500\,m$^2$ 
experimental area. Figure 1a shows the completed 
DRESDYN building from outside. Its 
left wing (LW) hosts a workshop, a chemistry lab, and a control room 
on the first floor. The central hall (CH) gives home to the liquid sodium 
experiments. The right wing (RW) contains most of the technical 
installations, in particular four sodium storage tanks for a total of 
12 tons, an electricity supply for up to 2.4\,MW, and a liquid argon fire 
extinguishing system with 15\,tons of liquid argon. The cleaning station 
(CS) in the foreground will serve for 
the final cleansing of sodium-spoiled installation parts with water.

\begin{figure}[ht!]
\includegraphics[width=0.95\textwidth]{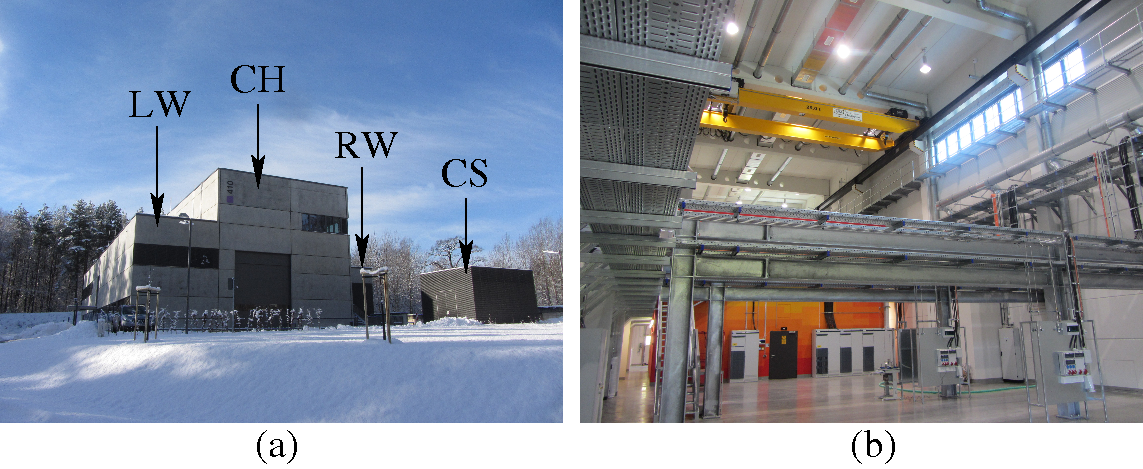}
\caption{The DRESDYN building at HZDR. (a) External view, 
(b) interior of the central hall.}
\label{fig:1}
\end{figure}

Figure 1b shows the interior of the central experimental hall. 
The reddish wall in the background is part of a special containment for the 
large-scale precession experiment. The two horizontal trusses in 
the foreground provide the sodium experiments with cooling media, 
electricity, and control data. Figure 2 illustrates 
the overall structure and the planned installations, 
with the separate containment for the precession experiments
(P), a large Taylor-Couette experiment (M) for investigations of the 
magnetorotational instability (MRI) and the Tayler instability, a 
sodium loop (L) and an ISI experiment (I). 

\begin{figure}[ht!]
\includegraphics[width=0.99\textwidth]{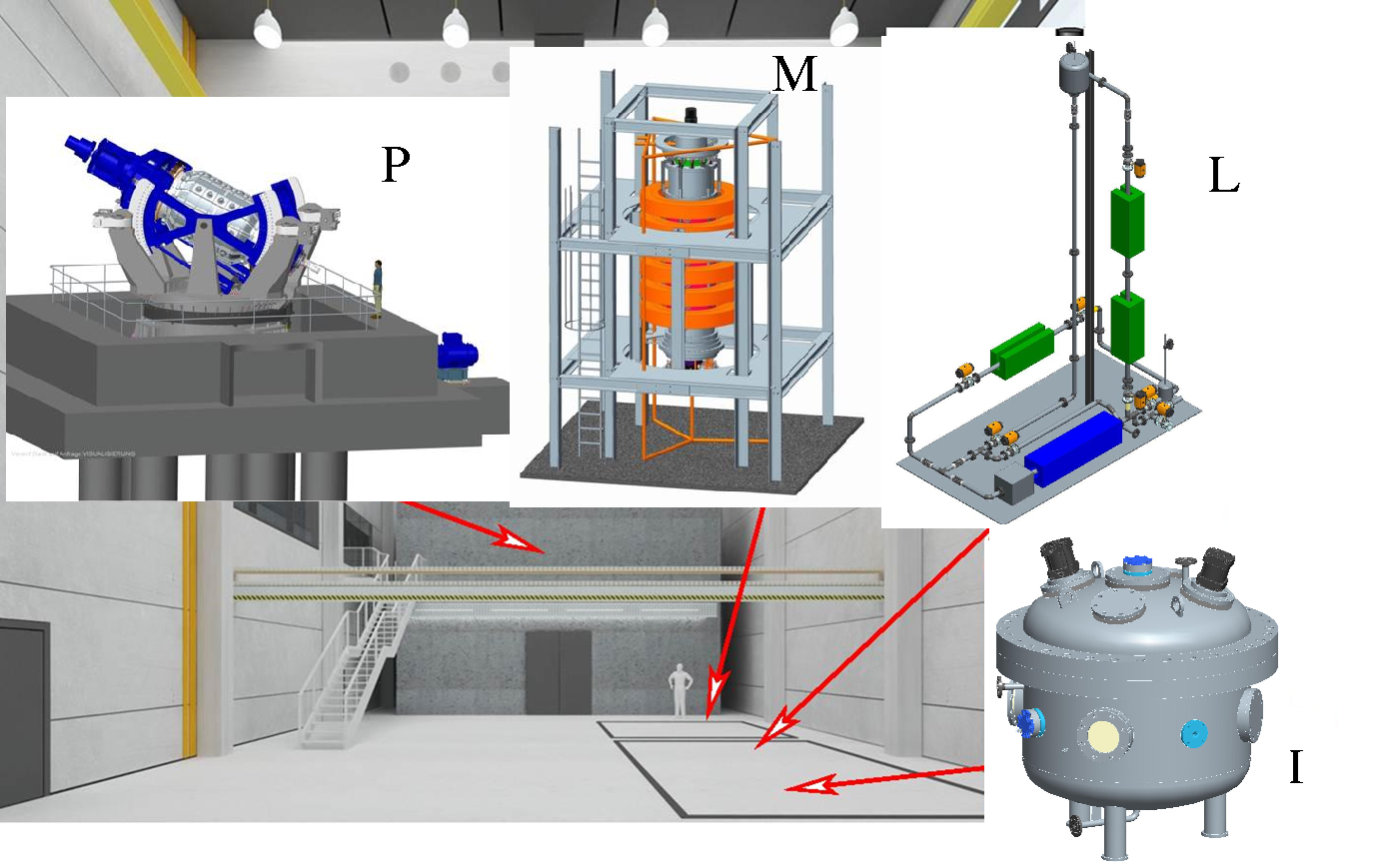}
\caption{Interior of the central hall, and the main planned 
experiments. Precession driven dynamo experiment (P) to be 
installed in the containment; Tayler-Couette experiment 
for the investigation of the magnetorotational and the 
Tayler instability (M); sodium loop (L); ISI 
experiment (I). A further stand for testing liquid 
metal batteries, which is not completely specified 
yet, will complement the list of sodium experiments.}
\label{fig:2}
\end{figure}

The most ambitious project in the framework of DRESDYN 
is certainly the large 
scale precession experiment (P) 
which is intended to continue the recent efforts 
in Riga
\cite{Gailitis2008}, Karlsruhe \cite{Mueller2004}, 
Cadarache \cite{Monchaux2009} 
and elsewhere to investigate homogeneous dynamo 
action in the liquid metal lab.

One of the key questions of geo- and astrophysical 
magnetohydrodynamics concerns the 
energy source of different 
cosmic dynamos. While thermal
and/or compositional buoyancy is considered 
the favourite candidate, 
precession
has also been discussed as a complementary energy 
source of the geodynamo
\cite{Malkus1968,Tilgner2005}, 
in particular at an early 
stage of Earth's evolution, prior to the
formation of the solid core. Some influence of 
orbital parameter variations can also
be inferred from paleomagnetic measurements that 
show an impact of the 100 kyr
Milankovic cycle of the Earth's orbit eccentricity 
on the reversal statistics of the
geomagnetic field \cite{Consolini2003,Stefani2006a}. 
Recently, precessional 
driving has also been discussed in
connection with the generation of the lunar 
magnetic field \cite{Dwyer2012}, and with dynamos
in asteroids \cite{Fu2013}.
In which parameters precession can drive a dynamo 
is still a matter of debate. While some simulations
in cylinders and cubes, carried out at 
magnetic Prandtl numbers $Pm \approx 1$, had found 
dynamo action for magnetic Reynolds numbers 
$Rm \approx 600$ \cite{Krauze2010,Nore2011,Stefani2015a},
more recent simulation at lower $Pm$ have pointed
to less optimistic values \cite{Giesecke2015,Goepfert2016}.

The DRESDYN precession experiment
consists of a cylindrical 
vessel of 2\,m diameter, 2\,m height, and 3\,cm wall 
thickness, supposed to 
rotate with 
up to 10\,Hz around its axis, and with up to 1\,Hz around a 
perpendicular axis. The mechanical and safety demands for such a 
large-scale sodium experiment are tremendous, in particular due to 
the huge gyroscopic torque (up to 8\,MNm) which requires a massive 
ferro-concrete basement which rests on 7 columns reaching 22\,m into 
the bedrock. 

A second experiment with  geo- and astrophysical background 
is a liquid sodium Taylor-Couette experiment (M) 
that is supposed to 
continue  and combine the recent experiments (using GaInSn) on 
the helical \cite{Stefani2006} and the azimuthal \cite{Seilmayer2014} 
versions of the magnetorotational instability (MRI) 
and on the Tayler 
instability (TI) \cite{Seilmayer2012}. Its main goal, however, is
to 
enhance the available parameter 
region in such a way that standard MRI \cite{Velikhov1959,Balbus1991}
should become reachable \cite{Ruediger2003}. 
Its central part is a sodium filled Tayler-Couette cell 
with a gap
width of 0.2 m and a height of 2 m, whose inner cylinder 
rotates with up to 20 Hz which corresponds to $Rm \approx 40$. 
Equipped with a coil system
for the generation of an axial field of up to 120 mT 
(giving a Lundquist number $S\approx 40$) and 
two different 
axial currents through the center and the liquid sodium,
this experiment aims at studying various versions of the 
magnetorotational instability and their combinations 
\cite{Kirillov2013} with  the Tayler instability
which might play an important role for 
the planetary synchronization of 
the solar dynamo \cite{Stefani2016}. 

The TI will also play a central role in a third experiment 
in which different
flow instabilities in liquid metal batteries (LMB) will 
be studied. LMBs consist of
three self-assembling liquid layers 
\cite{Kim2013}, an alkali or earth-alkali metal 
(Na, Mg), an
electrolyte, and a metal or half-metal 
(Bi, Sb). In order to be competitive, LMBs
have to be quite large, so that charging and 
discharging currents in the order of some kA 
are to be expected. Under those conditions, 
TI \cite{Weber2013,Weber2014}, electrovortex flows
\cite{Weber2015}, 
and interface instabilities \cite{Weber2017},
must be carefully avoided or at least controlled. 

Further to this, DRESDYN will also comprise a standard 
sodium loop (L in figure 2),
with a horizontal and a vertical test section 
of diameter 100\,mm. A 30 kW electromagnetic 
pump will provide a maximum flow rate of 56\,m$^3$/h. Both test 
sections will be equipped with various measurement flanges, 
and will optionally allow the application of magnetic fields 
in order to investigate particular MHD flow problems. The 
vertical test section will be used for 
investigations of two-phase flows of sodium and argon,
in particular using Mutual Induction Tomography 
\cite{Terzija2011}.

Another experimental test stand (I in figure 2) 
will be utilized for 
ISI experiments and tests of 
measurement techniques. 
Important flow measurement methods to be tested 
are the Ultrasonic Doppler Velocimetry (UDV)
\cite{Eckert2003,Eckert2011} and 
various inductive methods such as 
the phase-shift sensor \cite{Priede2011} and 
the transient 
eddy-current flow meter (TECFM) 
\cite{Forbriger2015}.
Besides of being calibration-free, TECFM 
is particularly interesting since the absence of 
any magnetic materials makes it 
suitable for high-temperature applications 
as they are relevant for SFRs.
In addition to these local methods, we will also
validate a  
global inductive method that aims at reconstructing entire 
velocity fields \cite{Stefani2004}. 
This Contactless Inductive Flow 
Tomography (CIFT) relies on the fact that externally applied 
magnetic fields are disturbed by the flow of a conducting medium. 
These small flow-induced modifications of the magnetic field are 
measured outside the liquid metal volume using an array of 
magnetic field sensors. By using this method, it is possible 
to reconstruct whole two-dimensional or three-dimensional 
velocity fields.

\section{Conclusion}

In this paper, we have presented the DRESDYN project 
as a platform for medium and large-scale liquid sodium 
experiments, partly with geo- and astrophysical motivation, 
partly related to energy technology problems. The 
sodium infrastructure is expected to 
become available by the 
end of 2017, so that first sodium experiments could 
start in 2018. Apart from the specific 
experiments discussed in this paper, DRESDYN is open 
for proposals of further liquid sodium investigation.
A large-scale Rayleigh-B\'enard experiment, 
potentially rotating and/or under
the influence of a magnetic field, might be 
a case in point.

\begin{acknowledgement}
Financial support of this research by the 
German Helmholtz Association in the frame of 
the Helmholtz-Alliance LIMTECH is gratefully 
acknowledged.
\end{acknowledgement}

\vspace{\baselineskip}

\end{document}